\documentclass{ws-p8-50x6-00}

\begin{document}

\newcommand\vev[1]{\langle{#1}\rangle}

\def\la{\mathrel{\mathpalette\fun <}}
\def\ga{\mathrelbe {\mathpalette\fun >}}
\def\fun#1#2{\lower3.6pt\vbox{\baselineskip0pt\lineskip.9pt
        \ialign{$\mathsurround=0pt#1\hfill##\hfil$\crcr#2\crcr\sim\crcr}}}

\renewcommand\({\left(}
\renewcommand\){\right)}
\renewcommand\[{\left[}
\renewcommand\]{\right]}

\newcommand\del{{\mbox {\boldmath $\nabla$}}}

\newcommand\eq[1]{Eq.~(\ref{#1})}
\newcommand\eqs[2]{Eqs.~(\ref{#1}) and (\ref{#2})}
\newcommand\eqss[3]{Eqs.~(\ref{#1}), (\ref{#2}) and (\ref{#3})}
\newcommand\eqsss[4]{Eqs.~(\ref{#1}), (\ref{#2}), (\ref{#3})
and (\ref{#4})}
\newcommand\eqssss[5]{Eqs.~(\ref{#1}), (\ref{#2}), (\ref{#3}),
(\ref{#4}) and (\ref{#5})}
\newcommand\eqst[2]{Eqs.~(\ref{#1})--(\ref{#2})}

\newcommand\pa{\partial}
\newcommand\pdif[2]{\frac{\pa #1}{\pa #2}}

\newcommand\yr{\,\mbox{yr}}
\newcommand\sunit{\,\mbox{s}}
\newcommand\munit{\,\mbox{m}}
\newcommand\wunit{\,\mbox{W}}
\newcommand\Kunit{\,\mbox{K}}
\newcommand\muK{\,\mu\mbox{K}}

\newcommand\metres{\,\mbox{meters}}
\newcommand\mm{\,\mbox{mm}}
\newcommand\cm{\,\mbox{cm}}
\newcommand\km{\,\mbox{km}}
\newcommand\kg{\,\mbox{kg}}
\newcommand\TeV{\,\mbox{TeV}}
\newcommand\GeV{\,\mbox{GeV}}
\newcommand\MeV{\,\mbox{MeV}}
\newcommand\keV{\,\mbox{keV}}
\newcommand\eV{\,\mbox{eV}}
\newcommand\Mpc{\,\mbox{Mpc}}

\newcommand\msun{M_\odot}
\newcommand\mpl{M_{\rm P}}
\newcommand\MPl{M_{\rm P}}
\newcommand\Mpl{M_{\rm P}}
\newcommand\mpltil{\widetilde M_{\rm P}}
\newcommand\mf{M_{\rm f}}
\newcommand\mc{M_{\rm c}}
\newcommand\mgut{M_{\rm GUT}}
\newcommand\mstr{M_{\rm str}}
\newcommand\mpsis{|m_\chi^2|}
\newcommand\etapsi{\eta_\chi}
\newcommand\luv{\Lambda_{\rm UV}}
\newcommand\lf{\Lambda_{\rm f}}

\newcommand\lsim{\mathrel{\rlap{\lower4pt\hbox{\hskip1pt$\sim$}}
    \raise1pt\hbox{$<$}}}
\newcommand\gsim{\mathrel{\rlap{\lower4pt\hbox{\hskip1pt$\sim$}}
    \raise1pt\hbox{$>$}}}

\newcommand\diff{\mbox d}

\def\dbibitem#1{\bibitem{#1}\hspace{1cm}#1\hspace{1cm}}
\newcommand{\dlabel}[1]{\label{#1} \ \ \ \ \ \ \ \ #1\ \ \ \ \ \ \ \ }
\def\dcite#1{[#1]}

\def\calm{{\cal M}}
\def\calp{{\cal P}}
\def\calr{{\cal R}}
\newcommand\calpr{\calp_\calr}                        

\newcommand\bfa{{\bf a}}
\newcommand\bfb{{\bf b}}
\newcommand\bfc{{\bf c}}
\newcommand\bfd{{\bf d}}
\newcommand\bfe{{\bf e}}
\newcommand\bff{{\bf f}}
\newcommand\bfg{{\bf g}}
\newcommand\bfh{{\bf h}}
\newcommand\bfi{{\bf i}}
\newcommand\bfj{{\bf j}}
\newcommand\bfk{{\bf k}}
\newcommand\bfl{{\bf l}}
\newcommand\bfm{{\bf m}}
\newcommand\bfn{{\bf n}}
\newcommand\bfo{{\bf o}}
\newcommand\bfp{{\bf p}}
\newcommand\bfq{{\bf q}}
\newcommand\bfr{{\bf r}}
\newcommand\bfs{{\bf s}}
\newcommand\bft{{\bf t}}
\newcommand\bfu{{\bf u}}
\newcommand\bfv{{\bf v}}
\newcommand\bfw{{\bf w}}
\newcommand\bfx{{\bf x}}
\newcommand\bfy{{\bf y}}
\newcommand\bfz{{\bf z}}

\newcommand\sub[1]{_{\rm #1}}
\newcommand\su[1]{^{\rm #1}}

\newcommand\supk{^{(K) }}
\newcommand\supf{^{(f) }}
\newcommand\supw{^{(W) }}
\newcommand\Tr{{\rm Tr}\,}

\newcommand\msinf{M\sub{inf}}
\newcommand\phicob{\phi\sub{COBE}}
\newcommand\mgrav{m_{3/2}(\phi)}
\newcommand\mgravsq{m_{3/2}^2(\phi)}
\newcommand\mgravcu{m_{3/2}^3(\phi)}
\newcommand\mgravvac{m_{3/2}}

\newcommand\pone{\dot\phi_1}
\newcommand\ptwo{\dot\phi_2}
\newcommand\ponesq{\dot\phi_1^2}
\newcommand\ptwosq{\dot\phi_2^2}
\newcommand\meff{m\sub{eff}}

\newcommand\cpeak{\sqrt{\tilde C_{\rm peak}}}
\newcommand\cpeako{\sqrt{\tilde C_{\rm peak}^{(0)}}}
\newcommand\omb{\Omega\sub b}
\newcommand\ncobe{N\sub{COBE}}

\newcommand\sigtil{\widetilde\sigma_8}
\newcommand\gamtil{\widetilde\Gamma}


\title{Self-tuning solutions of the cosmological constant  
\footnote
{Talk presented at SUSY'01, Dubna, Russia, 11-17 June 2001}
}
\author{Jihn E. Kim} 
\address{Department of Physics, Seoul National University, 
Seoul 151-747, Korea}

\maketitle

\abstracts{
I briefly review the cosmological constant problem and attempts toward
its solution, and present the first nontrivial example
for the self-tuning mechanism with a $1/H^2$ 
term with the antisymmetric field strength $H_{MNPQ}$
in a 5D RS-II setup. }

\section{Introduction}

It is generally believed that the cosmological constant
problem is the most severe hierarchy problem in particle
physics.\cite{ccreview} The hierarchy problem has been formulated
since 1975 in connection with grand unified theories(GUTs). GUTs
introduce two scales which differ by a factor of $10^{14}$. At the
classical Lagrangian level, there appear parameters which are of
order $10^{16}$ GeV. But loop corrections and spontaneous symmetry
breaking enter and after the dust settles down we require to have
an electroweak vacuum expectation value(VEV) of order 100~GeV. This
implies that the parameter of the Higgs boson mass must satisfy
$M_H^2+\Delta M_H^2=O(10^{-28})M^2_{GUT}$, which is a fine-tuning
problem known as the so-called gauge hierarchy problem. When we
consider different scales in the same Lagrangian, in general we 
encounter this kind of hierarchy problem.

Gravity is described by metric $g_{\mu\nu}$. Then the metric theory
of gravity is given by the action
\be
{\rm Action}=\int d^4x\sqrt{-g}\left(\frac{M^2}{2}R-V_0+\cdots\right)
\label{action}
\ee
where $V_0$ is the vacuum energy. The above action leads to
the Einstein equation
\be
R_{\mu\nu}-\frac{1}{2}R g_{\mu\nu}+8\pi G V_0 g_{\mu\nu}=8\pi G T_{\mu\nu}.
\ee
Actually, Einstein introduced the cosmological constant on the LHS
of the above equation as $\Lambda$ instead of $8\pi GV_0$ in
1917 to obtain a seemingly static universe. But the Hubble
expansion observed 12 years later invalidated this argument. Even
at the time of Einstein, the cosmological constant problem could
have been formulated as a hierarchy problem.
The parameter $M$ appearing in the action is the Planck mass
$M=2.44\times 10^{18}$~GeV. If gravity introduces this large mass,
then any other parameter in gravity is expected to be of that order,
in particular $V_0$ in Eq.~(\ref{action}). However, the bound on
the vacuum energy is very strong $<(0.01\ {\rm eV})^4$, which
implies a fine-tuning of order $10^{-120}$. Thus, this cosmological
constant problem is the most severe hierarchy problem.

Usually a hierarchy problem is understood if there exists a
symmetry related to it. The difficulty with the cosmological constant
problem is that there is no such symmetry working. One obvious
symmetry is the scale invariance but it is badly broken by
the mass terms, for example. The electroweak scale introduces a mass
scale which is about $10^{56}$ times larger than the observed
cosmological constant bound. The cosmological constant problem
has surfaced in particle physics for a need to set the minimum point
of the Higgs potential. But we cannot find a theoretical guideline 
where to put the vacuum energy, which is another way of stating the 
hierarchy problem.\cite{veltman}
Since then there have been several attempts toward a solution of the 
problem under the name of probabilistic interpretation,\cite{hawking} 
boundary of different phases,\cite{witten} wormholes,\cite{coleman} 
anthropic principle,\cite{weinberg} etc.\cite{etc}
Among these the most interesting ones are the probabilistic
interpretation and anthropic solution. 
Note that the probabilistic interpretation is based on the multi-vacua
possibility. 

The anthropic solution is
a working one as far as multi-vacua are allowed in the theory.
Probabilistic interpretation is also based on the multi-vacua
possibility. It is based on the requirement that life evolution
is not very much affected by the existence of the cosmological
constant. Galaxy formation may be hindered if the cosmological
constant is too large. Weinberg obtained a bound on the vacuum
energy density, $\rho<550\rho_c$, from the condition that 
condensation of matter occurs. Thus, in the anthropic
solution the fine-tuning is reduced to 1 out of 1000.

\section{Self-tuning Solutions
 }

\subsection{The old version}

For a given action, if there exists a flat space solution without 
a fine-tuning then it is called a self-tuning solution or an 
undetermined integration constant(UIC) solution in early eighties.
Note that in 4D it is not possible to have a flat space solution
$ds^2=d{\bf x}^2-dt^2$ with a nonvanishing $\Lambda$. Only de
Sitter space($\Lambda>0$) or anti de Sitter space($\Lambda<0$)
solutions are possible. Therefore, in 4D one needs an extreme
fine-tuning to satisfy the observed bound on the vacuum energy.

But suppose that there exists an UIC. Witten,\cite{witten}
using the earlier idea in the 11d supergravity,\cite{townsend} showed this
possibility with a four index antisymmetric field strength
$H_{\mu\nu\rho\sigma}$. In 4D, it is not a dynamical field but
equation of motion for $H_{\mu\nu\rho\sigma}$ can lead to a
constant contribution to the vacuum energy $\sim c^2$ via a nonvanishing
VEV $H_{\mu\nu\rho\sigma}=c\epsilon_{\mu\nu\rho\sigma}$. This UIC $c$
can be adjusted so that the final cosmological constant vanishes.
Once $c$ is determined there is no other parameter to adjust to
cancel further additions of the cosmological constant. Namely, at 
different stages of the spontaneous symmetry breaking(as shown in Table 1) 
we do not have
enough UIC. This example did not work because $H_{\mu\nu\rho\sigma}$
is not a dynamical field. Also, Hawking\cite{hawking} introduced
the four index antisymmetric field strength to explain his
probabilistic choice of vanishing cosmological constant in a multi
universe scenario. Again in 4D it is not a dynamical field and
there are not enough UIC.

In these old versions for the self-tuning solution, one {\it does not
require that only the flat space solution is the solution of
the equation of motion.} They allowed the de Sitter and anti de Sitter
space solutions. But the flat space is chosen from the other 
principles.\cite{hawking,witten}

\begin{table}[t]
\caption{Several fundamental scales accompanying vacuum energy
}
\begin{center}
\footnotesize
\begin{tabular}{|l|l|}
\hline
 (energy density)${}^\frac14$ & Physics of interest\\
\hline
$10^{18}\GeV$ & Gravity is strong \\
$10^{16}\GeV$? & Spontaneous breaking of GUT symmetry   \\
$10^{10-13}\GeV$? & PQ symmetry breaking \\
                  & Seed for SUSY breaking\\
$100\GeV$ &  Electroweak symmetry breaking \\
$1\GeV-100\MeV$ &  QCD chiral symmetry breaking \\
$10^{-3}\eV$ & Present vacuum energy \\
\hline
\end{tabular}
\end{center}
\label{table1}
\end{table}

\subsection{The new version} 

Recently, a more restriced class of self-tuning solutions 
is suggested for a solution of the cosmological constant 
problem.\cite{kachru} It requires the existence of a flat
space solution but forbid de Sitter and anti de Sitter
space solutions. It is a very fascinating idea, presumably
dreamed of for a long time. But the recent interest came
from the 5D brane scenario of Randall-Sundrum(RS).\cite{rs2}
The idea can be seen to be plausible by observing that
RS models can allow flat space solutions even starting with
nonvanishing brane tension and negative bulk cosmological
constant. However, toward the flat space solutions one should
have fine-tuned the parameters. Therefore, the RS models
can be a playground toward an effective 4D flat spacetime. 
In this spirit, the new version tried to
obtain the flat space solutions without fine-tuning
but without de Sitter and anti
de Sitter space solutions.\cite{kachru} However, their model
introduced an essential singularity and a proper treatment of
this singularity reintroduced a fine-tuning condition,\cite{nilles}
and hence it is fair to say that there has not appeared a
working self-tuning model in the new version. 

Since the rise and fall of this new version is interesting toward
future discovery of a self-tuning solution, I will briefly review it.

The RS-II model\cite{rs2} is an alternative to compactification.
With the bulk anti de Sitter space(AdS), the localization of gravity
makes it acceptable to have an uncompactified extra dimension.
A brane is located at $y=0$, where matter fields are assumed
to live. Then the 5D Lagrangian is
\begin{equation}
{\cal L}=\frac{M^3}{2}(R-\Lambda_b)+({\cal L}_{matter}-\Lambda_1)
\delta(y)
\end{equation}
where $M$ is a 5D Planck mass. One is interested in a 4D-flat solution,
\begin{equation}
ds^2=\beta(y)\eta_{\mu\nu}dx^\mu dx^\nu+dy^2.\label{metric}
\end{equation}
With a $Z_2$ symmetry, one can find a solution of (\ref{metric}),
$\beta(y)=\beta_0 \exp(-k|y|)$ where $k=(-\Lambda_b/6M^3)^{1/2}$.
The boundary condition at $y=0$ dictates a fine-tuning between
$k$ and $k_1=\Lambda_1/6$, $k=k_1$, where we set
$M=1$. If we put more branes, then there are more conditions to
satisfy due to the freedom to introduce more brane tension parameters
$\Lambda_i$ at brane $B_i$.
Thus, RS-II model is the easiest one to try for a self-tuning solution.
\\

\noindent \underline{\it First try}: Kachru et al. and Arkani-Hamed et 
al.\cite{kachru} tried the following 5D Lagrangian with a brane at $y=0$,
\begin{equation}
{\cal L}=R-\Lambda e^{a\phi}-\frac{4}{3}(\nabla\phi)^2
-Ve^{b\phi}\delta(y)
\end{equation}
where the fundamental scale $M/2$ is set to 1. It is a RS-II type model 
with a massless scalar field $\phi$ in the bulk. This scalar interacts
with the brane tension through the $V$ term. The relevant equations are
\begin{eqnarray}
diliaton &:&\ \ \ \frac{8}{3}\phi^{\prime\prime}+\frac{32}{3}A^\prime
\phi^\prime -a\Lambda e^{a\phi}-bV\delta(y)e^{b\phi}=0 \nonumber\\
(55) &:&\ \ \ 6(A^\prime)^2-\frac{2}{3}(\phi^\prime)^2+\frac{1}{2}\Lambda 
e^{a\phi}=0\\
(55),(\mu\nu) &:&\ \ \ 3A^{\prime\prime}+\frac{4}{3}(\phi^\prime)^2
+\frac{1}{2}e^{b\phi}V\delta(y) \nonumber
\end{eqnarray}
The 4D flat ansatz $A(y)=\ln\beta(y)^{1/2}$ may imply a zero
cosmological constant if there exists a solution of these equations.
Indeed, they found a bulk solution $\phi=\pm(3/4)\ln|(4/3)y+c|+d$,
with $A^\prime=\pm(1/3)\phi^\prime$ and $\Lambda=0$. Here, $c$ and
$d$ are integration constants and there remains only one integration
constant after satisfying the boundary condition at $y=0$. There are two
solutions, one without a singularity and the other with a singularity 
at $y_c=-(3/4)c$. The nonsingular solution diverges logarithmically
at large $|y|$ and localization of gravity near the brane is not
realized. The singular solution has a naked singularity at $y_c$.
An effective 4D theory is obtained by integrating out with $y$, and
hence we cannot ignore the space $y\ge y_c$. Even if the bulk for
a given $y$ is flat, the effective theory must know the whole
$y$ space. If we cut off the integration at 
$y=y_c-\epsilon$, then certainly the resultant 4D cosmological 
constant would not be zero, which is obvious since the singular 
point is an essential singularity. So the solution is incomplete
with the given flat space ansatz.
\\

\noindent\underline{\it Second try}: F$\ddot{\rm o}$rste et 
al.\cite{nilles} cured the singularity problem of the above example,
by inserting a brane at $y=y_c$. They showed that there exists a
solution but with one fine-tuning between parameters. 
Therefore, we have not obtained a new type self-tuning solution
yet.

\section{A Self-tuning Solution with $1/H^2$
 }

A naive try is to introduce a dynamical spin-0 field whose mass
is zero so that it affects the whole region of the bulk. Kachru
et al.\cite{kachru} in fact attempted to use a massless scalar.
In this case one must assume the form of the potential.
However, if a massless scalar such as a Goldstone boson arises 
from the symmetry in the theory, it will be much better.
It is achieved by an antisymmetric tensor field. For one
dynamical degree, the rank is $(2+n)$ in $(4+n)$-dimensional
spacetime. In 5D RS model, the antisymmetric tensor field is
$A_{MNP}(M,N,P=0,1,2,3,5)$ whose four-form field strength is
$H_{MNPQ}$. The four form field is invariant under the gauge
transformation,
\be
A_{MNP}\rightarrow A_{MNP}+
\partial_{[M}\lambda_{NP]}.
\ee
There will be a $U(1)$ gauge symmetry remaining with one massless
pseudoscalar field. The massless one is $a$: $\partial_Ma =
\sqrt{-g}\epsilon_{MNPQR}H^{NPQR}/4!$. Toward a flat-space solution
we adopt the following ansatze:
\be
Ansatz\ 1\ =\ (Flat\ 4D),\ \ 
Ansatz\ 2\ =\ (D=4\ chosen)\nonumber
\ee
where $\mu,\nu,\cdots$ are the 4D indices.\\

\centerline{\bf (i) With $H^2$ term}
The simplest term $H^2$
does not lead to a self-tuning solution. We use, 
\begin{eqnarray}
Ansatz\ &1&\ (Flat\ 4D)\ :\ ds^2=\beta(y)^2\eta_{\mu\nu}
dx^\mu dx^\nu+dy^2\nonumber \\
Ansatz\ &2&\ (D=4\ chosen)\ : H_{\mu\nu\rho\sigma}=
\epsilon_{\mu\nu\rho\sigma 5}\frac{1}{\sqrt{-g}n(y)}\nonumber
\end{eqnarray}
The field equation of $H$,
(55) and $(\mu\nu)$ component Einstein equations lead to the
following solutions 
\begin{eqnarray}
&\Lambda_b<0&:(a/k)^{1/4}[\sinh(4k|y|+c)]^{1/4}\nonumber\\
&\Lambda_b>0&:(a/k)^{1/4}[\sin(4k|y|+c')]^{1/4}\nonumber\\
&\Lambda_b=0&:(|4a|y|+c^{\prime\prime}|)^{1/4}\nonumber
\end{eqnarray}
For a localizable metric at $y=0$, there exists a singularity
at $-c/4k$, etc. Thus, another brane is necessary and
we need a fine-tuning as in the Kachru et al case.\cite{kachru}
\\

\centerline{\bf (ii) With $1/H^2$ term}
But we find that there exists a solution with $1/H^2$ term. 
Using the unit $M=1$, we take the following action,
\be
S=\int d^4x\int dy \sqrt{-g}\left\{
\frac{1}{2}R-\frac{2\cdot 4!}{H_{MNPQ}H^{MNPQ}}-\Lambda_b-\Lambda_1
\delta(y)
\right\}
\ee
As before we use the following ansatze,
\begin{eqnarray}
Ansatz\ &1&\ (Flat\ 4D)\ :\ ds^2=\beta(y)^2\eta_{\mu\nu}
dx^\mu dx^\nu+dy^2\nonumber \\
Ansatz\ &2&\ (D=4\ chosen)\ : H_{\mu\nu\rho\sigma}=
\epsilon_{\mu\nu\rho\sigma 5}\sqrt{-g}\frac{1}{n(y)}.\nonumber
\end{eqnarray}
The $H$ field equation is, $\partial_M[\sqrt{-g}H^{MNPQ}/H^4]=
\partial_\mu[\sqrt{-g}H^{\mu NPQ}/H^4]=0$. Therefore, $n$ is a
function of $y$ only. The (55) component Einstein equation is
$6(\beta'/\beta)^2=-\Lambda_b-(\beta^8/A)$, implying a solution 
in the anti de Sitter bulk $\Lambda_b<0$. Here, the integration
constant $A$ is positive and $2n^2=\beta^8/A$. 
The $(\mu\mu)$ equation is $3(\beta'/\beta)^2
+3(\beta^{\prime\prime}/\beta)=-\Lambda_b-\Lambda_1\delta(y)
-3(\beta^8/A)$. We impose a $Z_2$ symmetry. The boundary
condition at $y=0$ is $[\beta'/\beta]_{0^+}=-\Lambda_1/6$.
Then, the flat solution is found to be
\be
\beta(|y|)=\left[(a/k)\cosh(4k|y|+c)\right]^{-1/4}\label{i}
\ee
where $k=\sqrt{-\Lambda_b/6}, a=\sqrt{1/6A}$, and $k_1=\Lambda_1/6$.
The bulk solution (\ref{i}) contains integration constants $a$ and $c$.
Here, $a$ is basically the charge of the universe and is determined
by the definition of 4D Planck mass. But, $c$ is undetermined and 
fixed by the boundary condition at $y=0$,
\be
\tanh c=\frac{k_1}{k}=\frac{\Lambda_1}{\sqrt{-6\Lambda_b}}.\label{c}
\ee
Because of the limited range for $\tanh$, the solution is
possible in the region, $|\Lambda_1|<\sqrt{-\Lambda_b/6}$.
This relation shows that it is possible to have a flat space
solution for any value of the brane tension $\Lambda_1$ with anti
de Sitter space bulk. In this sense it is an old type self-tuning solution,
i.e. there exists a flat solution for any parameter value
within a finite range. If the observable sector VEV (i.e. $\Lambda_1$)
changes, the solution adjusts so that $c$ matches the condition
(\ref{c}). If there were no dynamical field, this adjustment would be
impossible, but in our case $H$ has one dynamical field and it is
possible to change the shape of the solution according to a changing
$\Lambda_1$.

Note that $\beta(|y|)$ is a decreasing function of $|y|$ and it
tends to 0 as $y\rightarrow \infty$. This property seems to be needed
for a self-tuning solution. The key points of our solution are:

\begin{itemize}
\item 
{\bf (1) $\beta$ has no naked singularity:}
If $\beta(y)$ has a singularity at $y=y_c$, we should cutoff the space
up to $y_c$. In this case, we would not obtain a vanishing 4D cosmological
constant. To avoid the singularity, we should introduce another brane
at $y=y_c$, and there results a fine-tuning since introduction of an 
additional brane introduces another brane tension which cannot be an
arbitrary parameter to give a flat space. Our solution
does not have a singularity in the whole $y$ space.
\item 
{\bf (2) Finite 4D Planck mass:}
Integrating with respect to $y$, we obtain a finite 4D Planck 
mass.\cite{kklcc}
\item  
{\bf (3) Self-tuning solution:}
We can show explicitly that by integrating out with $y$ the resultant 
4D cosmological constant is zero. In this consideration, 
the surface term gives an important contribution.\cite{kklcc}
\end{itemize}

So far we considered the time-independent solutions. The next
simpleset solutions are the de Sitter and anti de Sitter type
time dependent solutions. These are parametrized by the curvature
$\lambda$(+ for de Sitter and -- for anti de Sitter) with the
metric $g_{\mu\nu}=diag.(-1, e^{2\sqrt{\lambda}t}, 
e^{2\sqrt{\lambda}t}, e^{2\sqrt{\lambda}t})$ for dS$_4$ and 
$g_{\mu\nu}=diag.( -e^{2\sqrt{-\lambda}x_3}, 
e^{2\sqrt{-\lambda}x_3}, e^{2\sqrt{-\lambda}x_3}, 1)$ for AdS$_4$. 
The 4D Rieman tensor becomes $R_{\mu\nu}=3\lambda g_{\mu\nu}$
with $g_{\mu\nu}$ given above. One particularly interesting equation
is the (55) component equation
\be
(\beta')^2=\lambda -\frac{\lambda_b}{6}\beta^2-\frac{\beta^{10}}{6A}
.\label{desitter}
\ee
The cosmological constant obtained from these equations is $\lambda$.
In fact, in the RS-II model one can explicitly show this,\cite{kklcc}
using the exact solution given in Ref.~[11]. In our case, we have
not obtained an exact time dependent solution, but can show the
existence of de Sitter and anti de Sitter space solutions
theoretically and numerically.\cite{kklcc} For example the de Sitter
space solution is possible since
Eq.~(\ref{desitter}) allows a point $y_h$(the de Sitter space horizen) 
where $\beta'\ne 0$ for $\beta=0$. For the de Sitter space solution,
for example, integration from $-y_h$ to $+y_h$ should give
the vacuum energy $\lambda$ as in the RS-II case.\cite{kklcc} On the
other hand, the AdS$_4$ solution does not give a localized gravity.

\section{Conclusion and Comments
 }

We have obtained an old type self-tuning solution in the RS setup
using $1/H^2$ term: (i) from 5D, 4D is chosen by $\langle
H_{\mu\nu\rho\sigma}\rangle$, (ii) there is no tachyonic Kaluza-Klein
states, (iii) for some finite range of the brane tension there always 
exists a flat space solution, but (iv) we have not obtained an exact time
dependent solution.

Toward a solution of the cosmological constant problem, however, we note
two essential points encountered in our solution:
\begin{itemize}
\item
We adopted a peculiar kinetic energy term, $1/H^2$. In this sense the
solution can be considered as another fine-tuning. However, we
have shown that the flat space solution is from more general type of the
action $1/H^{2n} (n>0)$,\cite{kklcc} and hope that these forms will
be understood in the future.  A more immediate question is 
whether it gives consistent physics at low energy. One can consider a
consistent field theory model from ${\cal L}=
-(1/8)H^2(F^2)^2-(1/4)F^2$ where $F$ is a $U(1)$ field strength 
$F_{\mu\nu}$, i.e. $F^2=F_{\mu\nu}F^{\mu\nu}$. The Gaussian integral
would choose $F^2=-1/H^2$ and after integrating out the $U(1)$ field
we obtain $1/H^2$ term. Therefore, the consideration of $1/H^2$ as an
effective interaction below some energy scale makes sense. Since
$H$ corresponds to an exactly massless boson, it can be considered as
a Goldstone boson. Thus, we may construct a theory for a massless
Goldstone boson($\ne$ dilaton) toward a self-tuning solution.
\item
In our case, we observed that there exist de Sitter and anti de Sitter
space solutions also. This arose a question: Why the flat space?
In the literature, there already exists a proposal: the probabilistic
choice by Hawking.\cite{hawking} The probability to choose the
flat space with $\Lambda=0^+$ is exponentially larger than the other
cases. For this idea to work, the vacua allows multi universes
and it is so in our example. Duff's point\cite{duff} on the Hawking 
solution seems to have neglected the surface contribution. In our example,
a more concrete scenario would be possible if we obtain a time dependent
solution transforming one with a $\Lambda_1$ to another with a
different $\Lambda_1$. 
\end{itemize}

\vskip 0.5cm
\centerline{Acknowledgments}
This work is supported in part by the BK21 program of Ministry 
of Education, 
and by the Center for High Energy Physics(CHEP) of
Kyungpook National University. 



\end{document}